%% file: main.tex
\documentclass{article}
\usepackage{spconfa4,amsmath,graphicx}

\usepackage{booktabs}
\usepackage{multirow}
\usepackage{color}
\usepackage{bbding}

\usepackage[mode=buildnew]{standalone}
\usepackage{subfigure}

\usepackage{cite}

\ninept 

\title{Non-Causal to Causal SSL-Supported Transfer Learning: \\ Towards a High-Performance Low-Latency Speech Vocoder}
\name{Renzheng Shi$^{\ast}$, Andreas Bär$^{\ast}$, Marvin Sach$^{\ast}$, Wouter Tirry$^{\circ}$, Tim Fingscheidt$^{\ast}$}
\address{
$^{\ast}$Institute for Communications Technology, TU Braunschweig, Braunschweig, Germany \\
$^{\circ}$Goodix Technology (Belgium) BV, 3000 Leuven, Belgium
}
\begin{document}
%
\maketitle
\begin{abstract}

Recently, \texttt{BigVGAN} has emerged as high-performance speech vocoder. Its sequence-to-sequence-based synthesis, however, prohibits usage in low-latency conversational applications. Our work addresses this shortcoming in three steps. First, we introduce low latency into \texttt{BigVGAN} via implementing causal convolutions, yielding decreased performance. Second, to regain performance, we propose a teacher-student transfer learning scheme to distill the high-delay non-causal \texttt{BigVGAN} into our low-latency causal vocoder. Third, taking advantage of a self-supervised learning (SSL) model, in our case \texttt{wav2vec2.0}, we align its encoder speech representations extracted from our low-latency causal vocoder to the ground truth ones. In speaker-independent settings, both proposed training schemes notably elevate the performance of our low-latency vocoder, closing up to the original high-delay \texttt{BigVGAN}. At only 21\% higher complexity, our best small \textit{causal} vocoder achieves 3.96 PESQ and 1.25 MCD, excelling even the original small \textit{non-causal} \texttt{BigVGAN} (3.64 PESQ) by 0.32 PESQ and 0.1 MCD points, respectively.
\end{abstract}
\section{Introduction}
\label{sec:intro}

Speech vocoders aim to synthesize high-quality speech from acoustic features. Early successful neural vocoders, such as {\tt WaveNet} \cite{wavenet2016}, generate the speech waveform in an autoregressive manner, resulting in excessive inference time. To overcome this issue, non-autoregressive generative models have been studied, including flow \cite{waveglow2019, waveflow2020}, diffusion \cite{diffwave2021, wavegrad2021}, and generative adversarial network (GAN) models \cite{melgan2019, hifigan2020}. Out of the three options, GAN-based neural vocoders are favored due to higher inference speed and synthesis quality \cite{hifigan2020}.

A typical GAN setup, e.g., in {\tt MelGAN} \cite{melgan2019}, includes a fully convolutional generator with Mel spectrogram input and discriminators operating on time-domain speech signals at different scales. In {\tt HiFiGAN} \cite{hifigan2020}, the discriminator is extended by multi-period discriminators (MPDs) to achieve high-fidelity synthesis. Moreover, multi-resolution discriminators (MRDs) have been employed to enhance the spectrogram structure of the synthesized speech \cite{univnet2021}. Additionally, {\tt BigVGAN} \cite{bigvgan2023} introduces a learnable periodic composition module into the generator to incorporate periodic characteristics of the speech signal.

Despite the superiority of existing GAN-based neural vocoders, they operate in a sequence-to-sequence fashion which poses a limitation: By applying a convolution kernel along the time axis in a non-causal fashion, the prediction of the current frame is influenced by both past and potentially many future frames. Considering a conversational application, it is impractical to generate speech only after the speaker completes a sentence. Therefore, streamable GAN-based neural vocoders with \textit{causal} convolutions and with only a fixed small or even no lookahead have been applied for applications such as voice conversion \cite{streaming}, speech synthesis \cite{streamets2023}, speech enhancement \cite{streamenhancement2023}, and speech coding \cite{audiodec2023}.

Nonetheless, simply replacing non-causal convolutions with causal ones (thereby reducing the algorithmic delay) is expected to yield a performance degradation due to their insufficient capacity to represent the given data \cite{dualmodel2023}. A dual-mode architecture, enforcing shared weights between the causal and non-causal models, regains performance to some extent in voice conversion \cite{dualmodel2022}.
A less architecture-restrictive method is knowledge distillation \cite{kd2021}, where models with better modeling power are selected as the teacher \cite{ts2019}. Particularly for this case, in existing works \cite{ts2024, distilling2023}, the same model but with non-causal convolutions is adapted as the teacher to guide the causal student model. Meanwhile, to reduce the inconsistency between the non-causal and causal convolution, a training strate\-gy that employs two partially non-causal teacher models has been proposed \cite{tstraining2024}. However, they do not investigate the non-causal to causal transfer learning for a given, existing vocoder.

Recently, models trained with self-supervised learning (SSL) have demonstrated superior ability in extracting expressive representations of the input data. The extracted speech representations show a strong correlation to the acoustic or linguistic characteristics of the speech \cite{sssr2021}, facilitating various downstream tasks either by serving as an additional input condition \cite{sslhifigan2021, ssl_enhance2023} or loss function \cite{sssr_loss2023}. The question remains open whether SSL models can also be incorporated into a non-causal to causal transfer learning scheme.

In this paper, we propose a high-performance low-delay speech vocoder built upon {\tt BigVGAN} \cite{bigvgan2023}. First, we incorporate causal convolutions into the generator, yielding the expected performance drop. To regain performance while preserving the vocoder's causality, we then propose a non-causal to causal transfer learning scheme combined with supervision from self-supervised learned features. Taking account of the fact that vocoders are usually trained to generate time-domain speech from highly compressed speech representations (e.g., Mel spectrograms), we do not follow the typical knowledge distillation paradigm \cite{ts2024, distilling2023} which forces the causal student model to mimic the behavior of the non-causal teacher model. Instead, we perform {\it distillation via a feature matching loss inside the teacher discriminator} which was initially used in the adversarial training for the non-causal teacher vocoder, which poses a rather soft constraint on the student vocoder. Furthermore, inspired by SSL-based losses \cite{sssr_loss2023}, \texttt{wav2vec2.0} \cite{wav2vec_2.0} {\it representations are introduced} to further enhance the generalization ability of our causal student vocoder.

The rest of the paper is structured as follows. We describe our novel non-causal to causal SSL-supported transfer learning scheme for vocoders in Section 2. The employed experimental setup and results are presented in Section 3 and Section 4, respectively. Our work is concluded in Section 5.

\section{Proposed method}
\label{sec:method}

\subsection{Low-latency speech vocoder}
\label{sec:causal_model}

\begin{figure}[t]
  \centering
  \includegraphics[width=0.48\textwidth]{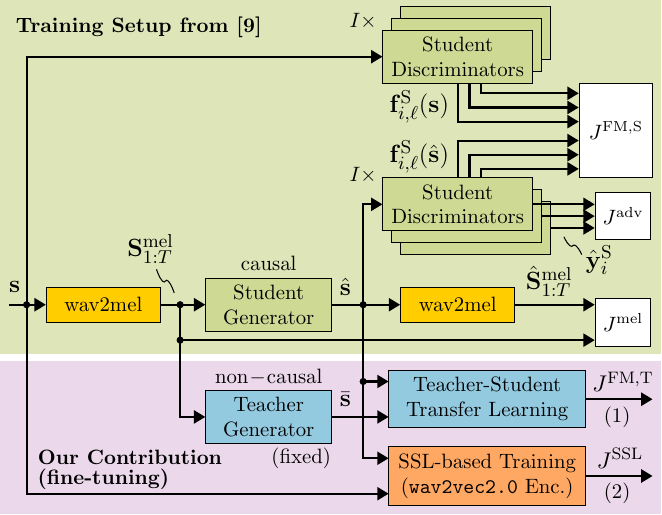}
  \caption{Proposed \textbf{non-causal to causal SSL-supported transfer learning}. We extend the default training setup (upper green part) by a teacher-student transfer learning module and an SSL-based training module (lower purple part, see Fig.\ \ref{fig:contribution} for further details). Details of the wav2mel block are shown in Fig.\ \ref{fig:wav2mel}.}
  \label{fig:training}
\end{figure}

Based on the (non-causal) {\tt BigVGAN} \cite{bigvgan2023}, we introduce causal convolutions to the generator to obtain a low-delay speech vocoder. This eventually means that all convolutions in our low-delay speech vocoder have no lookahead and solely rely on the current and past frames. We follow the original adversarial training protocol \cite{bigvgan2023} and visualize the training setup of our new {\it causal} vocoder (student generator) in Fig.\ \ref{fig:training} (upper green part). 

\begin{figure}[t]
	\centering
	\includegraphics[width=0.34\textwidth]{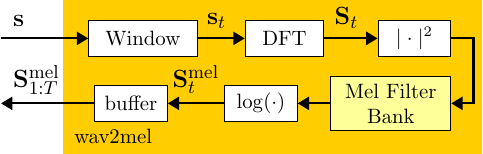}
	\caption{Details of the \textbf{wav2mel} block in Fig.\ \ref{fig:training}.}
  \label{fig:wav2mel}
\end{figure}

The ground truth speech waveform $\mathbf{s}$ is shown on the left. First, a wav2mel block, detailed in Fig.\ \ref{fig:wav2mel}, is applied to extract the sequence of Mel spectra $\mathbf{S}^{\mathrm{mel}}_{1:T}$. Specifically, the speech waveform $\mathbf{s}$ is divided into overlapping frames $\mathbf{s}_{t}$ by applying a periodic Hann window of length $N_{w}$ with a frame shift of $N_{s}$ samples. Here, $t \in \mathcal{T}$ denotes the frame index and $\mathcal{T}$ represents the set of frame indices, with $T=|\mathcal{T}|$ frames. The spectrogram $\mathbf{S}_{t}$ is acquired by applying the discrete Fourier transform (DFT) of length $K$ to $\mathbf{s}_{t}$, followed by an extraction of the squared amplitude spectrum. A Mel filter bank is utilized to obtain the respective Mel spectrum $\mathbf{S}^{\mathrm{mel}}_{t}$ with $M$ coefficients, which is logarithmically scaled and---in training---buffered. During training, the proposed {\it causal} vocoder then takes the Mel spectrogram $\mathbf{S}^{\mathrm{mel}}_{1:T}$ as input and outputs the synthesized speech waveform $\hat{\mathbf{s}}$, which is then followed by a second wav2mel block to obtain the Mel spectrogram $\hat{\mathbf{S}}^{\mathrm{mel}}_{1:T}$. During inference, the proposed {\it causal} vocoder takes the Mel spectra $\mathbf{S}^{\mathrm{mel}}_{t}$ as input and outputs the synthesized speech waveform frame-by-frame (low-latency).

Both the multi-period and multi-resolution discriminators from \texttt{BigVGAN} are employed (green student discriminator boxes in Fig.\ \ref{fig:training}). The discriminators take either the ground truth speech $\mathbf{s}$ or the synthesized speech $\hat{\mathbf{s}}$ as input. We define $\hat{\mathbf{y}}^{\mathrm{S}}_{i}$ as the $i$-th discriminator output, given synthesized speech $\hat{\mathbf{s}}$ input, and $\mathbf{f}_{i,\ell}^{\mathrm{S}}(\cdot)$ as (student discriminator) hidden states given, e.g., ground truth speech $\mathbf{s}$ or synthesized speech $\hat{\mathbf{s}}$ input. Here, indices $i \in \mathcal{I}$ and $\ell \in \mathcal{L}$ refer to the $i$-th student discriminator and its $\ell$-th hidden layer. In addition, sets $\mathcal{I}$ and $\mathcal{L}$ represent the set of discriminator indices and hidden layer indices, respectively, with $I = |\mathcal{I}|$ discriminators in total. We follow the \texttt{BigVGAN} training protocol and use three losses to train our \textit{causal} vocoder: adversarial loss $J^{\mathrm{adv}}$, L1 Mel spectrogram loss $J^{\mathrm{mel}}$, and feature matching loss $J^{\mathrm{FM, S}}$. For more details, please refer to the original {\tt BigVGAN} work \cite{bigvgan2023}.

\begin{figure}[t]
    \centering
    \subfigure[Teacher-student transfer learning]{
    \label{fig:teacher_student}
    \includegraphics[width=0.23\textwidth]{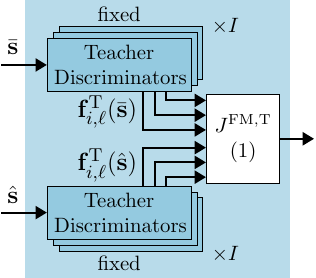}
    }
    \hfill
    \subfigure[SSL-based training]{
    \label{fig:lm_learning}
    \includegraphics[width=0.215\textwidth]{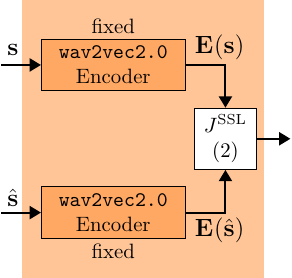}
    }
    \caption[toc entry]{Proposed (a) \textbf{teacher-student transfer learning} and (b) \textbf{SSL-based training} modules as used in Fig.\ \ref{fig:training}.}
    \label{fig:contribution}
\end{figure}

\subsection{Non-causal to causal transfer learning}
\label{sec:ts_framework}

Replacing non-causal convolutions with causal ones yields a low-latency speech vocoder, but also an expected performance drop. To address this issue, we propose a non-causal teacher to causal student transfer learning framework (see Fig.\ \ref{fig:training}, lower purple part, blue boxes).
In particular, a pre-trained {\tt BigVGAN} model with \textit{non-causal} convolutions employed in the generator serves as the teacher and our proposed {\it causal} vocoder as the student. The teacher shares the same discriminator setup as the student. Further details of the teacher-student transfer learning module are given in Fig.\ \ref{fig:teacher_student}.

Starting with Fig.\ \ref{fig:training}, the Mel spectrogram $\mathbf{S}_{1:T}^{\mathrm{mel}}$ is fed into the teacher generator to obtain the synthesized target speech waveform $\bar{\mathbf{s}}$. Continuing with Fig.\ \ref{fig:teacher_student}, the teacher discriminators then take either the synthesized speech $\bar{\mathbf{s}}$ from the teacher generator or $\hat{\mathbf{s}}$ from the student generator as input. Inspired by the GAN-based training \cite{melgan2019, hifigan2020, univnet2021, bigvgan2023}, we introduce a second feature matching loss
\begin{equation}
\label{eq:fmloss_t}
    J^{\mathrm{FM,T}}_{}=\frac{1}{|\mathcal{I}| \!\cdot\! |\mathcal{L}|} \sum\limits_{i\in\mathcal{I}} \sum\limits_{\ell\in\mathcal{L}} \! \big\|\mathbf{f}_{i,\ell}^{\mathrm{T}} (\bar{\mathbf{s}}) \!-\! \mathbf{f}_{i,\ell}^{\mathrm{T}} (\hat{\mathbf{s}})\big\|_{1},  
\end{equation}
where the L1 distance based on hidden layer outputs of the discriminators with inputs $\bar{\mathbf{s}}$ and $\hat{\mathbf{s}}$ is computed. Here, $\mathbf{f}_{i,\ell}^{\mathrm{T}}(\cdot)$ denotes the output of the $\ell$-th layer from the $i$-th \textit{teacher} discriminator. 

\subsection{Training with self-supervised learning (SSL) support}
\label{sec:ssl_loss}

We further introduce an SSL-supported training module (Fig.\ \ref{fig:training}, lower purple part, orange box) into our proposed non-causal to causal transfer learning framework with details shown in Fig.\ \ref{fig:lm_learning}. We use the encoder $\mathbf{E}()$ of the pre-trained {\tt wav2vec2.0} model \cite{wav2vec_2.0} to first extract speech representations for both, the ground truth input speech $\mathbf{s}$ and the student synthesized input speech $\hat{\mathbf{s}}$. Instead of a mean squared error (MSE) based loss as in \cite{sssr_loss2023}, we employ an SSL cosine similarity loss according to
\begin{equation}
\label{eq:ssl_loss}
    J^{\mathrm{SSL}}_{}(\mathbf{s}, \hat{\mathbf{s}})= 1 - \frac{\big(\mathbf{E}^{}_{}(\mathbf{s})\big)^{\mathrm{T}} \cdot \mathbf{E}^{}_{}(\hat{\mathbf{s}})}{\| \mathbf{E}^{}_{}(\mathbf{s}) \| \cdot \| \mathbf{E}^{}_{}(\hat{\mathbf{s}}) \|},  
\end{equation}
advocating similarity between the speech representations. The operator $(\cdot)^{\mathrm{T}}$ denotes the vector transpose.

\section{Experimental setup}
\label{sec:experiment}

\subsection{Dataset and preprocessing}
\label{sec:dataset}

We report results on the multi-speaker dataset VCTK \cite{VCTK} with all the recordings re-sampled to 16 kHz. Approximately 36.5 hours of recordings from 96 speakers are used as the training set $\mathcal{D}^{\text{train}}_{\text{VCTK}}$. Another 2.5 hours of speech with disjoint speakers are selected as the validation set $\mathcal{D}^{\text{val}}_{\text{VCTK}}$. Around one hour of speech from the remaining speakers are used to form the test set $\mathcal{D}^{\text{test}}_{\text{VCTK}}$.

The input Mel spectrogram features are obtained according to Fig.\ \ref{fig:wav2mel}. A Hann window of length $N_{w} = 512$ with a frame shift of $N_{s} = 128$ samples is employed. Further, a DFT of size $K=512$ and a set of $M=80$ Mel filters are used. For our proposed {\it causal} vocoder, the total algorithmic delay equals the window length, which is 32 ms for wideband speech.

\subsection{Model configurations}
\label{sec:training}

Two different generator designs are proposed in {\tt BigVGAN} \cite{bigvgan2023}. For a simple comparison, we use the stride setup of the upsampling layers as an additional label to distinguish them, i.e., the small {\tt BigVGAN} base is (8,8,2,2) and the large {\tt BigVGAN} is (4,4,2,2,2,2). For the training of the original small and large {\tt BigVGAN}, a window length of $N_{w}^{\prime} = 1024$ with a frame shift of $N_{s}^{\prime} = 256$ and a DFT of size $K=1024$ are used to obtain the Mel spectrogram features \cite{bigvgan2023}.

We first adapt the original {\tt BigVGAN} to the frame length $N_{w}$ and frame shift (FS) $N_{s}$ by changing the stride setup to (8,4,2,2) and (4,2,2,2,2,2), respectively. This new but still high-delay {\tt BigVGAN} is labeled as ``{\tt BigVGAN}, our FS''. Our proposed {\it causal} vocoder uses the same stride setup as ``{\tt BigVGAN}, our FS''. Note that all "our FS" methods, non-causal and causal ones, have the same size and complexity, the latter only determined by the (equal) kernel sizes.

\begin{table*}
    \centering
	\caption{Results of the reference \textbf{BigVGAN} \cite{bigvgan2023} with non-causal convolution (high-delay) and the proposed causal vocoder (low-delay) on the \textbf{VCTK test set} $\mathcal{D}^{\text{test}}_{\text{VCTK}}$ at 16 kHz. Both the large and small model setups are investigated. Our frame shift (``our FS'') indicates a shorter frame length and frame shift. The \textbf{best} and \underline{second-best} results among our proposed large and small low-delay vocoders  (32 ms) are highlighted.}
	\input{tables/causal_ablation_all1}
    \label{tab:causal_ablation_all}
\end{table*}

\subsection{Training setup and evaluation}
\label{sec:metric}

We build upon the setup of {\tt BigVGAN} \cite{bigvgan2023} using the official implementation to train all models. Further training details are as follows.

\textbf{Pre-training (Stage 1).} As shown in Fig.\ \ref{fig:training}, the causal vocoder (denoted as the student generator) is trained with the adversarial loss $J^{\mathrm{adv}}$, the feature matching loss $J^{\mathrm{FM,S}}$, and the Mel spectrogram loss $J^{\mathrm{mel}}$ for 1M steps with an initial learning rate of $10^{-4}$. Please refer to the {\tt BigVGAN} \cite{bigvgan2023} for further information.

\textbf{Fine-tuning (Stage 2).} First, the teacher vocoder, i.e., the non-causal {\tt BigVGAN} model that shares the same stride setup with the student vocoder, is trained with 1M steps using the same training setup shown in Fig.\ \ref{fig:training}. Then, a fine-tuning of the student model with the proposed SSL-supported teacher-student transfer learning module is performed using a learning rate of $3 \cdot 10^{-4}$. Again, all the losses from the first stage are employed along with the proposed second feature matching loss (\ref{eq:fmloss_t}) and the SSL loss (\ref{eq:ssl_loss}), see Fig.\ \ref{fig:contribution}, resulting in the final loss
\begin{equation}
\label{eq:genloss_ft}
\begin{aligned}
    J^{\mathrm{gen}}_{} = J^{\mathrm{adv}} &+ \lambda^{\mathrm{mel}} \cdot J^{\mathrm{mel}} + \lambda^{\mathrm{FM}} \cdot J^{\mathrm{FM,S}} \\ &+ \lambda^{\mathrm{FM}} \cdot J^{\mathrm{FM,T}} + \lambda^{\mathrm{SSL}} \cdot J^{\mathrm{SSL}},
\end{aligned}
\end{equation}
with hyperparameters $\lambda^{\mathrm{mel}}=45$,  $\lambda^{\mathrm{FM}}=2$, and $\lambda^{\mathrm{SSL}}=4$. The student vocoder is fine-tuned for another 1M steps.

\textbf{Quality metrics.} Three instrumental measures are employed for evaluation, namely, the perceptual evaluation of speech quality (PESQ) according to ITU-T Recommendation P.826.2 \cite{PESQ}, the Mel-cepstral distance (MCD) \cite{MCD}, and the phone similarity score PSS$\,$=$\,$100$\,$-$\,$LPD$\,$(\%), based on the Levenshtein phoneme distance (LPD) \cite{lpd2023}.
It reports, in a language-independent fashion, the similarity of international phonetic alphabet (IPA) phones recognized on the synthesized speech $\hat{\mathbf{s}}$ as compared to the ground truth speech $\mathbf{s}$.

\section{Results and discussion}
\label{sec:result}

In Table \ref{tab:causal_ablation_all}, we compare our proposed low-delay {\it causal} vocoder against three models on the VCTK test set $\mathcal{D}^{\text{test}}_{\text{VCTK}}$: the high-delay {\tt BigVGAN} \cite{bigvgan2023} (baseline), our new high-delay ``{\tt BigVGAN}, our FS'', and the {\tt BigVGAN} with causal convolutions (causal {\tt BigVGAN}). Following Section \ref{sec:training}, we investigate two generator designs, i.e., small and large, starting with the large one in the upper table segment and the small one in the lower table segment. Model stride setup and its respective algorithmic delay are complemented by the model complexity, measured in terms of trainable parameter count (\# Params.) and the number of floating-point operations per second (\# GFLOPS). Finally, all quality measures from Section \ref{sec:metric}, i.e., PESQ, MCD, and PSS = 100 - LPD (\%), are reported.

\textbf{Large causal low-delay vocoder.} First, in the upper table segment of Table \ref{tab:causal_ablation_all}, for the \textit{large} generator design, we observe that our new ``{\tt BigVGAN}, our FS'' significantly outperforms the baseline {\tt BigVGAN} in all quality measures, e.g., PESQ improves from 4.17 to 4.44. This shows that the proposed smaller frame length and frame shift (see Section \ref{sec:training}) is beneficial for speech quality. Due to the shorter frame shift, this comes with a somewhat higher computational complexity (152.23 vs.\ 114.60 GFLOPS). Incorporating causal convolutions into the {\tt BigVGAN} (causal {\tt BigVGAN}) yields a limited algorithmic delay of 64 ms, however, at the price of much lower speech quality compared to the baseline {\tt BigVGAN} and ``{\tt BigVGAN}, our FS''. Now, taking advantage of a smaller frame length and frame shift, our proposed large {\it causal} vocoder outperforms the causal {\tt BigVGAN} by 0.17 PESQ points and 0.08 MCD points. By only applying the proposed non-causal to causal transfer learning scheme (T/S training), the performance further improves by 0.08 PSEQ points (3.93 vs.\ 3.85), 0.07 MCD points (1.32 vs.\ 1.39), and particularly PSS rises from 97.35\% to 97.73\%. By only incorporating the SSL training from Sec.\ \ref{sec:ssl_loss}, we observe a significant improvement of up to 0.18 PESQ points (4.03 vs.\ 3.85) and 0.17 MCD points (1.23 vs.\ 1.39). Finally, the \textit{best performance} for our proposed large {\it causal} vocoder is achieved by \textit{using the proposed non-causal to causal SSL-supported transfer learning scheme}. Compared with the baseline {\tt BigVGAN}, \textit{we recovered much 
of the speech quality}.

\textbf{Small causal low-delay vocoder.} Do our proposed methods transfer to the \textit{small} generator? To answer this, we look at the lower segment of Table \ref{tab:causal_ablation_all}. We again observe the profit of using a smaller frame length and frame shift with our new ``{\tt BigVGAN} base, our FS'' compared to the baseline {\tt BigVGAN} base model. This time, the gap is even larger, improving the PESQ from 3.64 to a stunning 4.26, surpassing even the baseline (large) {\tt BigVGAN} performance (4.17 PESQ, first row). Similar to before, a naive utilization of causal convolutions (causal {\tt BigVGAN} base) is worst across all quality measures. \textit{On the contrary, our ``proposed small {\it causal} vocoder, our FS'' already shows comparable performance to the {\tt BigVGAN} base model with an algorithmic delay of just 32 ms}. In addition, with the non-causal to causal transfer learning scheme, we reach an improvement of 0.13 PESQ points (3.79 vs.\ 3.67) and 0.1 MCD points (1.30 vs.\ 1.40). Employing the SSL training instead gives significant improvements of 0.27 PESQ points (3.94 vs.\ 3.67) and 0.11 MCD points (1.29 vs.\ 1.40). By combining both---along with "our FS"--- \textit{our best small {\it causal} vocoder achieves a PESQ of 3.96, MCD of 1.25, and PSS of 97.79\%, excelling the baseline {\tt BigVGAN} base model by 0.32 PESQ points, 0.1 MCD points, and 1\% absolute PSS} at only 21\% higher complexity (47.76 vs. 39.46 GFLOPS). 

\textbf{Ablation on T/S training and SSL training}. In Table \ref{tab:ab_finetune}, we show ablation studies on the formulation of the SSL loss $J^{\mathrm{SSL}}$, the SSL model inputs, and the teacher discriminator inputs for the proposed transfer learning scheme, carried out with our small {\it causal} vocoder from Table \ref{tab:causal_ablation_all}. We first look at SSL loss (\ref{eq:ssl_loss}) alternatives, namely the mean squared error (MSE) \cite{sssr_loss2023} and the mean absolute error (MAE) between speech representations. We then conduct experiments on inputs of the SSL model to compute (\ref{eq:ssl_loss}) and on the teacher discriminators to find the best target for the student to learn from. Throughout these ablations, the stride setup is (8,4,2,2) and the results are reported on the VCTK \textit{validation} set $\mathcal{D}^{\text{val}}_{\text{VCTK}}$.

First, we observe that our vocoder trained with the SSL loss formulated with cosine similarity (\ref{eq:ssl_loss}) achieves the best PESQ (3.94) and second-best MCD (1.28). For the inputs of the proposed learning schemes, we observe that our vocoder benefits more from mimicking the SSL representations extracted from the ground truth speech $\mathbf{s}$, excelling the one using the teacher output $\bar{\mathbf{s}}$ by 0.02 PESQ points and 0.02 MCD points. Looking at the proposed non-causal to causal transfer learning scheme, our vocoder gives slightly better results by using the teacher output $\bar{\mathbf{s}}$ as the target (0.02 PESQ and 0.01 MCD points). Since the teacher generator is trained to synthesize speech such that the teacher discriminator cannot distinguish it from the ground truth speech, the hidden states from the teacher discriminators exhibit a higher correlation compared to the one from the ground truth speech. On the other hand, the SSL model is trained to extract expressive speech representations from the ground truth speech. Thus, using the ground truth speech as the target in our proposed SSL training scheme gives better guidance.

\begin{table}	
    \vspace*{-2mm}
    \centering
	\caption{Ablation study of the proposed teacher-student training and SSL loss on $\mathcal{D}^{\text{val}}_{\text{VCTK}}$. All vocoders are causal and use the proposed (8,4,2,2) stride setup. \textbf{Best} and \underline{second-best} results are highlighted.}
	\input{tables/ablation_finetune1}
    \label{tab:ab_finetune}
\end{table}

\section{Conclusions}
\label{sec:conclusion}

We proposed a low-latency speech vocoder from {\tt BigVGAN} and a novel non-causal to causal transfer learning scheme to improve its performance. We show that the causal student vocoder benefits from 
the non-causal teacher discriminator. Further, a self-supervised learning (SSL) model is integrated to enhance the causal student vocoder in modeling spectral relations. Putting all together, along with different frame length and shift, we obtain a low-latency (student) vocoder that achieves a PESQ of 3.96 and MCD of 1.25, improving the original non-causal {\tt BigVGAN} (PESQ of 3.64) by an impressive 0.32 PESQ points, 0.1 MCD points, and 1\% absolute phone similarity score (96.78\% vs. 97.79\%), respectively.

\clearpage
\bibliographystyle{IEEEbib}
\bibliography{refs}

\end{document}

%% file: tables/causal_ablation_all1.tex
\setlength{\tabcolsep}{2mm}{
\begin{tabular*}{17.05cm}{lc cccccc}

	\toprule



    \textbf{Model} & \textbf{Stride Setup} & \textbf{Alg. Delay} &\textbf{\# Params.} & \textbf{\# GFLOPS} & \textbf{PESQ}$\big\uparrow$ & \textbf{MCD}$\big\downarrow$ & \textbf{PSS}$\big\uparrow$ \\

    \midrule
    {\tt BigVGAN} \cite{bigvgan2023} & (4,4,2,2,2,2) & \multirow{2}{*}{high} & 112.23 M & 114.60 & 4.17 & 0.86 & 97.91$\,$\% \\
	{\tt BigVGAN}, our FS & (4,2,2,2,2,2) & & 111.05 M & 152.23 & 4.44 & 0.63 & 98.30$\,$\% \\
    \midrule
    causal {\tt BigVGAN} & (4,4,2,2,2,2) & 64 ms & 112.23 M & 114.60 & 3.68 & 1.47 & 96.99$\,$\% \\
    \midrule
	proposed large causal vocoder, our FS & (4,2,2,2,2,2) & 32 ms & 111.05 M & 152.23 & 3.85 & 1.39 & 97.35$\,$\% \\
    $\;$ w/ T/S training & (4,2,2,2,2,2) & 32 ms & 111.05 M & 152.23 & 3.93 & 1.32 & 97.73$\,$\% \\
    $\;$ w/ SSL training & (4,2,2,2,2,2) & 32 ms & 111.05 M & 152.23 & \underline{4.03} & \underline{1.23} & \textbf{97.85$\,$\%} \\
    $\;$ w/ T/S \& SSL training & (4,2,2,2,2,2) & 32 ms & 111.05 M & 152.23 & \textbf{4.05} &  \textbf{1.21} & \underline{97.83$\,$\%} \\
 
    \midrule
    {\tt BigVGAN} base \cite{bigvgan2023} & (8,8,2,2) & \multirow{2}{*}{high} & 13.95 M & 39.46 & 3.64 & 1.35 & 96.78$\,$\% \\
	{\tt BigVGAN} base, our FS & (8,4,2,2) & & 13.69 M & 47.76 & 4.26 & 0.90 & 98.06$\,$\% \\
    \midrule
    causal {\tt BigVGAN} base & (8,8,2,2) & 64 ms & 13.95 M & 39.46 & 3.13 & 1.98 & 95.45$\,$\% \\
    \midrule
	proposed small causal vocoder, our FS & (8,4,2,2) & 32 ms & 13.69 M & 47.76 & 3.67 & 1.40 & 97.15$\,$\% \\
    $\;$ w/ T/S training & (8,4,2,2) & 32 ms & 13.69 M & 47.76 & 3.79 & 1.30 & 97.38$\,$\% \\
    $\;$ w/ SSL training & (8,4,2,2) & 32 ms & 13.69 M & 47.76 & \underline{3.94} & \underline{1.29} & \underline{97.64$\,$\%} \\
    $\;$ w/ T/S \& SSL training & (8,4,2,2) & 32 ms & 13.69 M & 47.76 & \textbf{3.96} & \textbf{1.25} & \textbf{97.79$\,$\%} \\

	\bottomrule
\end{tabular*}}

%% file: tables/ablation_finetune1.tex
\setlength{\tabcolsep}{2mm}{
\begin{tabular*}{8.2cm}{l cc}

	\toprule


    \textbf{Model} & \textbf{PESQ}$\big\uparrow$ & \textbf{MCD}$\big\downarrow$ \\
    

    \midrule
    proposed small causal vocoder, our FS & 3.67 & 1.40 \\
    \midrule
    w/ $J^{\mathrm{SSL}}$ in Fig. 2 being $\ldots$ &  & \\
    $\ldots$ eq. (2) & \textbf{3.94} & \underline{1.28} \\
    $\ldots$ MSE \cite{sssr_loss2023} & \underline{3.92} & \textbf{1.27} \\
    $\ldots$ MAE & \underline{3.92} & 1.32 \\
    \midrule
    w/ $J^{\mathrm{SSL}}$ (2) based on $\ldots$ & & \\
    $\ldots$ ground truth $\mathbf{s}$, student output $\hat{\mathbf{s}}$ & \textbf{3.94} & \textbf{1.28} \\
    $\ldots$ teacher output $\bar{\mathbf{s}}$, student output $\hat{\mathbf{s}}$ & 3.92 & 1.26 \\
    \midrule
    w/ discriminator inputs for T/S training: &  & \\
    $\;$ ground truth $\mathbf{s}$, student output $\hat{\mathbf{s}}$ & 3.81 & 1.32 \\
    $\;$ teacher output $\bar{\mathbf{s}}$, student output $\hat{\mathbf{s}}$ & \textbf{3.83} & \textbf{1.31} \\

	\bottomrule
\end{tabular*}}

%% file: main.bbl
\begin{thebibliography}{10}

\bibitem{wavenet2016}
A.~van~den Oord, S.~Dieleman, H.~Zen, K.~Simonyan, O.~Vinyals, A.~Graves, N.~Kalchbrenner, A.~Senior, and K.~Kavukcuoglu,
\newblock ``{WaveNet: A Generative Model for Raw Audio},''
\newblock {\em arXiv preprint arXiv:1609.03499}, Sept. 2016.

\bibitem{waveglow2019}
R.~Prenger, R.~Valle, and B.~Catanzaro,
\newblock ``{WaveGlow: A Flow-Based Generative Network for Speech Synthesis},''
\newblock in {\em Proc. of ICASSP}, Brighton, UK, May 2019, pp. 3617--3621.

\bibitem{waveflow2020}
W.~Ping, K.~Peng, K.~Zhao, and Z.~Song,
\newblock ``{WaveFlow: A Compact Flow-Based Model for Raw Audio},''
\newblock in {\em Proc. of ICML}, Virtual, July 2020, pp. 7706--7716.

\bibitem{diffwave2021}
Z.~Kong, W.~Ping, J.~Huang, K.~Zhao, and B.~Catanzaro,
\newblock ``{Diff\-Wave: A Versatile Diffusion Model for Audio Synthesis},''
\newblock in {\em Proc. of ICLR}, virtual, May 2021, pp. 1--17.

\bibitem{wavegrad2021}
N.~Chen, Y.~Zhang, H.~Zen, R.~J. Weiss, M.~Norouzi, and W.~Chan,
\newblock ``{WaveGrad: Estimating Gradients for Waveform Generation},''
\newblock in {\em Proc. of ICLR}, virtual, May 2021, pp. 1--15.

\bibitem{melgan2019}
K.~Kumar, R.~Kumar, T.~de~Boissiere, L.~Gestin, W.~Z. Teoh, J.~Sotelo, A.~de~Brébisson, Y.~Bengio, and A.~Courville,
\newblock ``{MelGAN: Generative Adversarial Networks for Conditional Waveform Synthesis},''
\newblock in {\em Proc. of NeurIPS}, Vancouver, BC, Canada, Dec. 2019, pp. 14910--14921.

\bibitem{hifigan2020}
J.~Kong, J.~Kim, and J.~Bae,
\newblock ``{HiFi-GAN: Generative Adversarial Networks for Efficient and High Fidelity Speech Synthesis},''
\newblock in {\em Proc. of NeurIPS}, Vancouver, BC, Canada, Dec. 2020, pp. 17022--17033.

\bibitem{univnet2021}
W.~Jang, D.~Lim, J.~Yoon, B.~Kim, and J.~Kim,
\newblock ``{UnivNet: A Neural Vocoder with Multi-Resolution Spectrogram Discriminators for High-Fidelity Waveform Generation},''
\newblock in {\em Proc. of Interspeech}, Brno, Czechia, Aug. 2021, pp. 2207--2211.

\bibitem{bigvgan2023}
S.~Lee, W.~Ping, B.~Ginsburg, B.~Catanzaro, and S.~Yoon,
\newblock ``{BigVGAN: A Universal Neural Vocoder With Large-Scale Training},''
\newblock in {\em Proc. of ICLR}, Kigali, Rwanda, May 2023, pp. 1--20.

\bibitem{streaming}
Z.~Chen, H.~Miao, and P.~Zhang,
\newblock ``{Streaming Non-Autoregressive Model for Any-to-Many Voice Conversion},''
\newblock {\em arXiv preprint arXiv:2206.07288}, June 2022.

\bibitem{streamets2023}
K.~Scheck, D.~Ivucic, Z.~Ren, and T.~Schultz,
\newblock ``{Stream-ETS: Low-Latency End-to-End Speech Synthesis from Electromyography Signals},''
\newblock in {\em Proc. of ITG Speech Communication}, Aachen, Germany, Sept. 2023, pp. 200--204.

\bibitem{streamenhancement2023}
K.~Kobayashi, T.~Hayashi, and T.~Toda,
\newblock ``{Low-Latency Electrolaryngeal Speech Enhancement Based on FastSpeech2-Based Voice Conversion and Self-Supervised Speech Representation},''
\newblock in {\em Proc. of ICASSP}, Rhodes, Greece, June 2023, pp. 1--5.

\bibitem{audiodec2023}
Y.~Wu, I.~D. Gebru, D.~Markovic, and A.~Richard,
\newblock ``{Audiodec: An Open-Source Streaming High-Fidelity Neural Audio Codec},''
\newblock in {\em Proc. of ICASSP}, Rhodes, Greece, June 2023, pp. 1--5.

\bibitem{dualmodel2023}
Z.~Ning, Y.~Jiang, P.~Zhu, J.~Yao, S.~Wang, L.~Xei, and M.~Bi,
\newblock ``{DualVC: Dual-Mode Voice Conversion Using Intra-Model Knowledge Distillation and Hybrid Predictive Coding},''
\newblock in {\em Proc. of Interspeech}, Dublin, Ireland, Aug. 2023, pp. 2063--2067.

\bibitem{dualmodel2022}
T.~Hayashi, K.~Kobayashi, and T.~Toda,
\newblock ``{An Investigation of Streaming Non-Autoregressive Sequence-to-Sequence Voice Conversion},''
\newblock in {\em Proc. of ICASSP}, Singapore, Singapore, May 2022, pp. 6802--6806.

\bibitem{kd2021}
S.~Stanton, P.~Izmailov, P.~Kirichenko, A.~A. Alemi, and A.~G. Wilson,
\newblock ``{Does Knowledge Distillation Really Work?},''
\newblock in {\em Proc. of NeurIPS}, Virtual, Dec. 2021, pp. 6906--6919.

\bibitem{ts2019}
R.~Aihara, T.~Hanazawa, Y.~Okato, G.~Wichern, and J.~L. Roux,
\newblock ``{Teacher-Student Deep Clustering for Low-Delay Sing\-le Channel Speech Separation},''
\newblock in {\em Proc. of ICASSP}, Brighton, United Kingdom, Apr. 2019, pp. 690--694.

\bibitem{ts2024}
{Y. Ai and Z. Ling},
\newblock ``{Low-Latency Neural Speech Phase Prediction based on Parallel Estimation Architecture and Anti-Wrapping Losses for Speech Generation Tasks},''
\newblock {\em IEEE/ACM T-ASLP (early access)}, pp. 1--14, Apr. 2024.

\bibitem{distilling2023}
K.~Tanaka, H.~Kameoka, T.~Kaneko, and S.~Seki,
\newblock ``{Distilling Sequence-to-Sequence Voice Conversion Models for Streaming Conversion Applications},''
\newblock in {\em Proc. of SLT}, Doha, Qatar, Jan. 2023, pp. 1022--1028.

\bibitem{tstraining2024}
K.~Wakayama, T.~Ochiai, M.~Delcroix, M.~Yasuda, S.~Saito, S.~Araki, and A.~Nakayama,
\newblock ``{Self-Supervised Learning for Speech Enhancement Through Synthesis},''
\newblock in {\em Proc. of ICASSP}, Seoul, Korea, Apr. 2024, pp. 561--565.

\bibitem{sssr2021}
Y.~Chung, Y.~Belinkov, and J.~Glass,
\newblock ``{Similarity Analysis of Self-Supervised Speech Representations},''
\newblock in {\em Proc. of ICASSP}, Toronto, ON, Canada, June 2021, pp. 3040--3044.

\bibitem{sslhifigan2021}
A.~Polyak, Y.~Adi, J.~Copet, E.~Kharitonov, K.~Lakhotia, W.~Hsu, A.~Mohamed, and E.~Dupoux,
\newblock ``{Speech Resynthesis from Discrete Disentangled Self-Supervised Representations},''
\newblock in {\em Proc. of Interspeech}, Brno, Czechia, Aug. 2021, pp. 3615--3619.

\bibitem{ssl_enhance2023}
B.~Irvin, M.~Stamenovic, M.~Kegler, and L.~Yang,
\newblock ``{Self-Supervised Learning for Speech Enhancement Through Synthesis},''
\newblock in {\em Proc. of ICASSP}, Rhodes, Greece, June 2023, pp. 1--5.

\bibitem{sssr_loss2023}
G.~Close, W.~Ravenscroft, T.~Hain, and S.~Goetze,
\newblock ``{Perceive and Predict: Self-Supervised Speech Representation Based Loss Functions for Speech Enhancement},''
\newblock in {\em Proc. of ICASSP}, Rhodes, Greece, June 2023, pp. 1--5.

\bibitem{wav2vec_2.0}
A.~Baevski, H.~Zhou, A.~Mohamed, and M.~Auli,
\newblock ``{Wav2vec 2.0: A Framework for Self-Supervised Learning of Speech Representations},''
\newblock in {\em Proc. of NeurIPS}, Vancouver, Canada, Dec. 2020, pp. 12449--12460.

\bibitem{VCTK}
C.~Veaux, J.~Yamagishi, and K.~MacDonald,
\newblock ``{CSTR VCTK Corpus: English Multi-Speaker Corpus for CSTR Voice Cloning Toolkit},''
\newblock {\em {University of Edinburgh. The Centre for Speech Technology Research (CSTR)}}, 2017.

\bibitem{PESQ}
{I}{T}{U},
\newblock {\em {Rec. P.862.2: Corrigendum 1, Wideband Extension to Recommendation P.862 for the Assessment of Wideband Telephone Networks and Speech Codecs}},
\newblock {International Telecommunication Standardization Sector (ITU-T)}, Oct. 2017.

\bibitem{MCD}
R.~F. Kubichek,
\newblock ``{Mel-Cepstral Distance Measure for Objective Speech Quality Assessment},''
\newblock in {\em Proc. of PACRIM}, Victoria, BC, Canada, May 1993, pp. 125--128.

\bibitem{lpd2023}
J.~Pirklbauer, M.~Sach, K.~Fluyt, W.~Tirry, W.~Wardah, S.~Möller, and T.~Fingscheidt,
\newblock ``{Evaluation Metrics for Generative Speech Enhancement Methods: Issues and Perspectives},''
\newblock in {\em Proc. of ITG Speech Communication}, Aachen, Germany, Sept. 2023, pp. 265--269.

\end{thebibliography}
